\documentclass[longauth, traditabstract]{aa} 
\usepackage{graphicx}
\usepackage{txfonts}
\usepackage{natbib}
\begin{document}

\title{Herschel observations of extra-ordinary sources:  Detecting spiral arm clouds by CH absorption lines\thanks {Herschel is an ESA space
    observatory with science instruments provided by European-led
    Principal Investigator consortia and with important participation
    from NASA.}}

\author{
S.-L. Qin,\inst{1} P. Schilke,\inst{1,2} C. Comito,\inst{2} T.~
M\"oller,\inst{1} R. Rolffs,\inst{2} H.~S.~P. M\"uller,\inst{1} A.
Belloche,\inst{2} K.~M. Menten,\inst{2} D.~C.~Lis,\inst{3}
T.~G.~Phillips,\inst{3} E.~ A. Bergin,\inst{4}
T.~A.~Bell,\inst{3} N.~R.~Crockett,\inst{4}
G.~A. Blake,\inst{3} S. Cabrit,\inst{5} 
E.~Caux,\inst{6,7} C.~Ceccarelli,\inst{8}
J.~Cernicharo,\inst{9} F.~Daniel,\inst{9,10}
M.-L.~Dubernet,\inst{11,12} M.~Emprechtinger,\inst{3}
P.~Encrenaz,\inst{10} E.~Falgarone,\inst{10} 
M.~Gerin,\inst{10} T.~F.~Giesen,\inst{1}
J.~R.~Goicoechea,\inst{9} P.~F.~Goldsmith,\inst{13}  H.~Gupta,\inst{13}
E.~Herbst,\inst{14} C.~Joblin,\inst{6,7} D.~Johnstone,\inst{15}
W.~D. Langer,\inst{13} S.~D.~Lord,\inst{16} S.~Maret,\inst{8}
P.~G.~Martin,\inst{17} G.~J.~Melnick,\inst{18} P.~Morris,\inst{13}
J.~A.~Murphy,\inst{19} D.~A.~Neufeld,\inst{20}
V.~Ossenkopf,\inst{1,21} L. Pagani,\inst{5} J.~C.~Pearson,\inst{13}
M.~P\'erault,\inst{10} R.~Plume,\inst{22} M. Salez,\inst{5} S.~Schlemmer,\inst{1}
J.~Stutzki,\inst{1} N.~Trappe,\inst{19} F.~F.~S.~van der
Tak,\inst{21} C.~Vastel,\inst{6,7} S.~Wang,\inst{4}
H.~W.~Yorke,\inst{13} S.~Yu,\inst{13} J.~Zmuidzinas,\inst{3}
A.~Boogert,\inst{16} R.~G\"usten,\inst{2} P.~Hartogh,\inst{23}
N.~Honingh,\inst{1} A.~Karpov,\inst{3} J.~Kooi,\inst{3}
J.-M.~Krieg,\inst{10} R.~Schieder, \inst{1} M.~C.~Diez-Gonzalez,\inst{24}
R.~Bachiller,\inst{24} J.~Martin-Pintado,\inst{9} W.~Baechtold,\inst{25}
M.~Olberg,\inst{26} L.~H.~Nordh,\inst{27} J.~L.~Gill,\inst{13} \and
G.~Chattopadhyay\inst{13}
}

\institute{I. Physikalisches Institut, Universit\"at zu K\"oln,
              Z\"ulpicher Str. 77, 50937 K\"oln, Germany \\
             \email{qin@ph1.uni-koeln.de}
\and  Max-Planck-Institut f\"ur Radioastronomie, Auf dem H\"ugel 69, 53121 Bonn, Germany
\and California Institute of Technology, Cahill Center for Astronomy and Astrophysics 301-17, Pasadena, CA 91125 USA
\and Department of Astronomy, University of Michigan, 500 Church Street, Ann Arbor, MI 48109, USA
\and LERMA \& UMR8112 du CNRS, Observatoire de Paris, 61, Av. de l'Observatoire, 75014 Paris, France
\and Centre d'\'etude Spatiale des Rayonnements, Universit\'e de Toulouse [UPS], 31062 Toulouse Cedex 9, France
\and CNRS/INSU, UMR 5187, 9 avenue du Colonel Roche, 31028 Toulouse Cedex 4, France
\and Laboratoire d'Astrophysique de l'Observatoire de Grenoble,
BP 53, 38041 Grenoble, Cedex 9, France.
\and Centro de Astrobiolog\'ia (CSIC/INTA), Laboratiorio de Astrof\'isica Molecular, Ctra. de Torrej\'on a Ajalvir, km 4
28850, Torrej\'on de Ardoz, Madrid, Spain
\and  LERMA, CNRS UMR8112, Observatoire de Paris and \'Ecole Normale Sup\'erieure, 24 Rue Lhomond, 75231 Paris Cedex 05, France
\and LPMAA, UMR7092, Universit\'e Pierre et Marie Curie,  Paris, France
\and LUTH, UMR8102, Observatoire de Paris, Meudon, France
\and Jet Propulsion Laboratory,  Caltech, Pasadena, CA 91109, USA
\and Departments of Physics, Astronomy and Chemistry, Ohio State University, Columbus, OH 43210, USA
\and National Research Council Canada, Herzberg Institute of Astrophysics, 5071 West Saanich Road, Victoria, BC V9E 2E7, Canada
\and Infrared Processing and Analysis Center, California Institute of Technology, MS 100-22, Pasadena, CA 91125
\and Canadian Institute for Theoretical Astrophysics, University of Toronto, 60 St George St, Toronto, ON M5S 3H8, Canada
\and Harvard-Smithsonian Center for Astrophysics, 60 Garden Street, Cambridge MA 02138, USA
\and  National University of Ireland Maynooth. Ireland
\and  Department of Physics and Astronomy, Johns Hopkins University, 3400 North Charles Street, Baltimor
e, MD 21218, USA
\and SRON Netherlands Institute for Space Research, PO Box 800, 9700 AV, Groningen, The Netherlands
\and Department of Physics and Astronomy, University of Calgary, 2500
University Drive NW, Calgary, AB T2N 1N4, Canada
\and MPI f\"ur Sonnensystemforschung, D 37191 Katlenburg-Lindau,
Germany
\and Observatorio Astron\'omico Nacional (IGN), Centro Astron\'omico de Yebes, Apartado 148. 19080 Guadalajara,  Spain
\and Microwave Laboratory, ETH Zurich, 8092 Zurich, Switzerland
\and Chalmers University of Technology, SE-412 96 Göteborg, Sweden, Sweden/ SRON Netherlands Institute for Space Research, Landleven 12, 9747 AD Groningen, the Netherlands
\and Department of Astronomy, Stockholm University, SE-106 91 Stockholm, Sweden
}


\abstract{We have observed CH absorption lines ($J=3/2, N=1
\leftarrow J=1/2, N=1$) against the continuum source Sgr~B2(M)
using the \textit{Herschel}/HIFI instrument. With the high
spectral resolution and wide velocity coverage provided by HIFI,
31 CH absorption features with different radial velocities and
line widths are detected and identified. The narrower line width
and lower column density clouds show `spiral arm' cloud
characteristics, while the absorption component with the broadest line
width and highest column density corresponds to the gas from the
Sgr~B2 envelope. The observations show that each `spiral arm'
harbors multiple velocity components, indicating that the clouds
are not uniform and that they have internal structure. This line-of-sight
through almost the entire Galaxy offers unique possibilities to
study the basic chemistry of simple molecules in diffuse clouds,
as a variety of different cloud classes are sampled
simultaneously. We find that the linear relationship between CH
and H$_2$ column densities found at lower $A_V$ by UV observations
does not continue into the range of higher visual extinction.
There, the curve flattens, which probably means that CH is
depleted in the denser cores of these clouds.}

   \titlerunning{Spiral arm clouds by CH absorption lines}
    \authorrunning{S.-L. Qin et al.}
   \maketitle
%

\section{Introduction}

While physical and chemical conditions of molecular clouds
associated with star-forming regions are widely explored by
observing various molecular species, much less is known about the
chemistry of the bulk of molecular clouds in the spiral arms of
our Galaxy. These have mostly been studied only in the CO
molecule, with some exceptions. These exceptions consist of clouds
that happen to lie along the line-of-sight to a bright Galactic or
extragalactic continuum source, and can be observed in absorption.
Examples for spiral arm clouds probed by gas absorption are the
Sgr~B2, W49, W51, and Cas~A millimeter continuum sources, where the
chemistry was studied mostly through low-energy lines of molecular
species
(e.g., Greaves \& Williams 1994; Greaves \& Nyman 1996; Gerin et
al. 2010; Menten et al. 2010; Lis et al. 2010; 
Schilke et al. 2010; Tieftrunk et al. 1994; Wyrowski et al. 2010).
Liszt and Lucas (2002) also made important contributions to
this field through observations of absorption line clouds against
quasars. The observations show that most of the absorption signal
arises from diffuse/translucent clouds with low gas density and
excitation temperature, suggesting that the ambient ultraviolet
field plays a prominent role in determining the chemistry (Lis et
al. 2010; Gerin et al. 2010; Neufeld et al. 2010).

Simple linear molecules are fundamental constituents when studying
interstellar chemistry, since they are building blocks for more complex
species (e.g., Liszt 2009). The very light hydride radical CH
(methylidyne) has been observed in diffuse clouds in front of
bright OB stars at optical wavelengths and in dark clouds at radio
wavelengths, showing that CH column density is linearly correlated
with the optical extinction (A$_{v}$) for A$_{v}$ less than
4$^{m}$ (e.g., Lang \& Wilson 1978; Federman 1982; Sheffer et al.
2008).
Observations and chemical
models also indicate that CH abundance is lower by a factor of 100-1000
in dense clouds and is mainly located in the surface layers of
dense clouds (Viala, Bel \& Clavelet 1979; de Jong, Genzel et al. 1979; Boland \&
Dalgarno 1980; Goicoechea, Rodriguez-Fern\'andez \& Cernicharo
2004; Polehampton et al. 2007) or low column density clouds
(Mattila 1986; Magnani, Lugo \& Dame 2005).

The prominent Sagittarius B2 star-forming region is very close
($\sim$ 130 pc) to the Galactic center and at a distance of $\sim$8
kpc from the Sun (Reid et al. 2009).
The dense cores contained in it, Sgr~B2(N) and Sgr~B2(M), are well-studied
massive star-forming regions.
Sgr~B2(M) is stronger at submm wavelengths and also less
contaminated by emission lines than its neighbor Sgr~B2(N)
(Goldsmith et al. 1990; Nummelin et al. 1998), making it 
the target of choice for absorption studies.
HIFI, the Heterodyne
Instrument for the Far-Infrared (de Graauw et al. 2010) onboard
the {\it Herschel} Space Observatory (Pilbratt et al. 2010) is an
ideal instrument for making such observations. In this Letter, we
present high spectral-resolution HIFI observations of CH towards
Sgr~B2(M).

\section{Observations}
A full spectral scan of Sgr~B2(M) in the HIFI band 1a, covering a
frequency range from 479.6 to 560.3 GHz was carried out on March 1, 2010,
 and was pointed at $\alpha$(J2000) =
17$^{h}$47$^{m}$20.2$^{s}$ and $\delta$(J2000)
=$-$28$^{\circ}$23$^{\prime}$05.0$^{\prime\prime}$, using the dual
beam switch (DBS) mode as part of the guaranteed time key program
HEXOS: {\it Herschel/HIFI Observations of Extraordinary sources:
The Orion and Sagittarius B2 Star forming Regions} (Bergin et al. 2010).
The DBS reference beams lie approximately 3$^{\prime}$ east and west (i.e.
perpendicular to the roughly north-south elongation of Sgr~B2).
The 1.1 MHz spectral resolution of the Wide Band Spectrometer (WBS)
 corresponds to a velocity
resolution of $\sim$0.6 km~s$^{-1}$ over a 4 GHz IF bandwidth. The
data have been processed through the standard pipeline released
with version 2.9 of HIPE (Ott et al. 2010), including removal of
the standing waves, and subsequently exported to FITS format. The
double-sideband (DSB) spectra were deconvolved into
single-sideband spectra, including the continuum (Comito \&
Schilke 2002) using the IRAM GILDAS package. The systematic
velocity of the source was corrected to the local stand of rest
(LSR) frame. Both H and V polarization data were obtained. The
data presented here are averages of the two polarizations with
equal weighting. The HIFI beam size at 532 GHz is
$\sim$37$^{\prime\prime}$, with a main beam efficiency of 0.69.


\section{Results}
The low-frequency radio transitions of CH are between $\Lambda$-type doublet
states, while spin-rotational as well as rotational transitions fall in the
far-infrared and submillimeter bands (Brown \& Evenson 1983). Early
observations mainly focused on the low-frequency radio and far-infrared
lines, the latter with relative low spectral resolution. The lowest
spin-rotational transitions of CH at 532.8 and 536.8 GHz were
measured in the laboratory by Amano (2000),
thus lying in a range of strong atmospheric water absorption,
which makes their observation with ground-based telescope impossible.
The level populations for the 532.8 and 536.8 GHz transitions are
expected to have a low excitation temperature in most clouds, with the possible exception of the Sgr~B2 envelope, and thus provide a
good measure of the total column density from absorption measurement.

Six hyperfine components of CH in the ground electronic state
($X$ $^{2}$$\prod$), $J=3/2, N=1 \leftarrow J=1/2, N=1$
near 532.8 ($1+\leftarrow1-, 2+\leftarrow1-$, and
$1+\leftarrow0-$), and 536.8 GHz $(2-\leftarrow1+,
1-\leftarrow1+$, and $1-\leftarrow0+)$ are included in the HIFI
band 1a observations of Sgr~B2(M).
The CH spectra near 532.8 and
536.8 GHz obtained with HIFI are presented in Fig. 1, with
multiple velocity components seen in absorption. The observations
of CH at $J=3/2, N=2 \leftarrow J=1/2, N=2$ in
 far-infrared ($\sim$ 2010 GHz) with a velocity resolution of $>$30
km~s$^{-1}$ towards Sgr~B2 using the {\it Infrared Space
Observatory (ISO)} and the Kuiper Airborne Observatory (KAO) 91.4
cm telescope showed absorption centered on 65 km~s$^{-1}$ with an
unusually large line width of $\sim$200 km~s$^{-1}$ (Stacey,
Lugten \& Genzel 1987; Goicoechea, Rodriguez-Fern\'andez \&
Cernicharo 2004; Polehampton et al. 2007), which samples the
gas from both the Galactic center and spiral arm clouds along the
light of sight.
The HIFI CH spectra are spectrally resolved into multiple velocity
components with different line intensities and widths. The strong
CH absorption at $\sim$64 km~s$^{-1}$ with the largest line width
clearly corresponds to the envelope of the Sgr~B2(M) cloud. In
addition to the Sgr~B2(M) absorption cloud, the radial velocities
of the other absorption components are consistent with those of
the spiral arm clouds and kinematic features associated with the
Galactic center (e.g., Greaves \& Williams 1994; Menten et al.
2010). HIFI observations of H$_{2}$O$^{+}$ and H$_{2}$O in the
spiral arm clouds against the Sgr~B2(M) continuum show somewhat
different spectral profiles than CH (Schilke et al. 2010; Lis et
al. 2010). Probably each molecular species samples slightly
different regions within various clouds along the line of sight.

%
%
%
%
\begin{figure}[h]
\centering
\includegraphics[width=10cm,height=9cm,angle=-90]{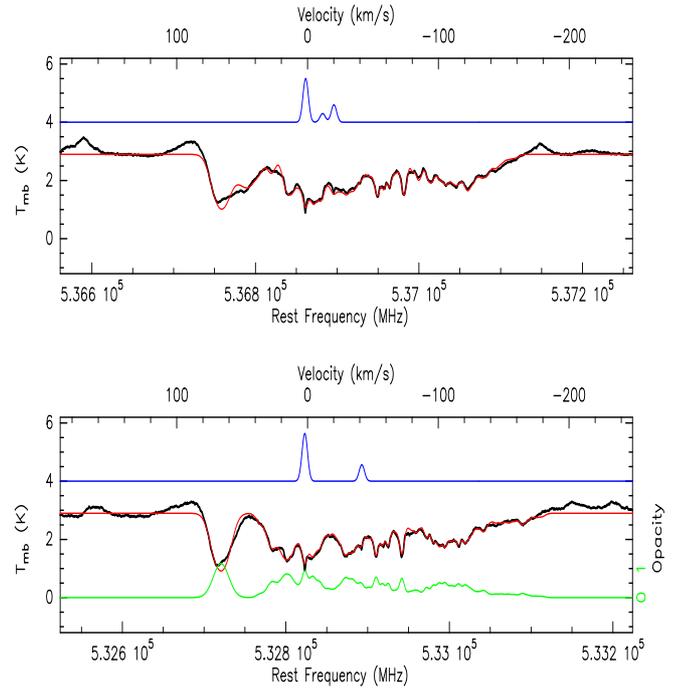}

\caption{CH spectra with velocity scale referring to the hyperfine
lines at 536.761145 and 532.723926 GHz. The upper panel shows the
spectrum at 536.8 GHz, the lower panel shows the spectrum at 532.8 GHz.
The green curve in the bottom panel represents the total opacity as a function of  LSR velocity, obtained by
linearly adding up the opacities of the hyperfine components. In each panel, the
black curve is the observed spectrum, the LTE fit is shown as the
red curve, and the blue curve shows the hyperfine pattern for the transition. }\label{}
\end{figure}

Absorption-line column densities can be calculated from the
observed line-to-continuum ratio by Gaussian fitting to the lines,
if the absorption spectrum comes from a single molecular-species 
without hyperfine components, as long as the
spectrum is not contaminated by emission from other species. We
have not identified any strong interfering lines in the velocity
range of the 532.8 and 536.8 GHz CH absorption. However, the hyperfine structure of CH makes it difficult to uniquely
assign the absorption to distinct velocity components and
calculate their column densities.
Taking the hyperfine components and
contamination from other species into account, and using the LTE
approximation, we modeled the CH spectra by employing XCLASS\footnote{We
made use of the myXCLASS program
(https://www.astro.uni-koeln.de/projects/schilke/XCLASS), which
accesses the CDMS (M\"uller at al. 2001, M\"uller et al. 2005;
http://www.cdms.de) and JPL (Pickett et al. 1998;
http://spec.jpl.nasa.gov) molecular data bases.}
and the automated fitting routine provided by
MAGIX\footnote{https://www.astro.uni-koeln.de/projects/schilke/MAGIX}.
We adopted 2.73 K as rotation temperature for all absorption
components, which is questionable only for the Sgr~B2 component itself,
since the envelope is hot and dense enough to excite CH.  A reliable estimate of the CH column density and abundance in this component has to await detailed radiative transfer modeling.
Keeping the velocity offsets fixed, we used 31
absorption components and 1 emission component, as for
c-C$_{3}$H$_{2}$ and H$^{13}$CO$^{+}$ cases (Menten et al. 2010)
for the CH model fitting. We assume that the opacity at any
velocity is a sum of contributions from all CH hyperfine components and that the absorbing layers cover the continuum completely.
The best fits to the spectra along with the line opacity as a function of velocity are shown in Fig. 1.
The column densities and line widths of the 31 absorption
components of the best-fit model are given in Table 1. Based on
the results of Menten et al. (2010) and Greaves \& Williams
(1994), the corresponding name of the spiral arm clouds for each
velocity component is also given in Table 1.
 \begin{table}
\caption{Model-fitting results and abundances}
\begin{center}
 \begin{tabular}{ccccc}
 \hline\hline
 $V_{\mathrm{lsr}}$ & $\Delta V$ & $N_{\rm CH}$&$N_{\rm H_{2}}^a$& $f_{\rm CH}$ \\
 (km~s$^{-1}$) & (km~s$^{-1}$) & (cm$^{-2}$)& (cm$^{-2}$) \\
\hline

64.0$^b$ & 13 & 5.2$\times10^{14}$&5.4$\times10^{23}$ & 9.6$\times10^{-10}$ \\
38.5$^c$ & 6  & 2.3$\times10^{13}$&1.9$\times10^{21}$ & 1.2$\times10^{-8}$ \\
32.0$^c$ & 7 & 6.4$\times10^{13}$&6.3$\times10^{21}$ & 1.0$\times10^{-8}$ \\
22.0$^d$ & 7 & 1.1$\times10^{14}$&3.0$\times10^{22}$ & 3.8$\times10^{-9}$  \\
17.0$^d$ & 5 & 7.8$\times10^{13}$&2.5$\times10^{22}$ & 3.2$\times10^{-9}$ \\
12.0$^d$ & 5 & 6.1$\times10^{13}$&2.2$\times10^{22}$ & 2.8$\times10^{-8}$ \\
8.2$^e$  & 4 & 1.3 $\times10^{14}$&2.4$\times10^{21}$ & 5.4$\times10^{-8}$ \\
2.5$^e$  & 5 & 1.2 $\times10^{14}$&1.2$\times10^{22}$ & 1.0$\times10^{-8}$ \\
-2.5$^e$ & 5 & 9.2$\times10^{13}$&4.8$\times10^{21}$& 1.9$\times10^{-8}$  \\
-8.7$^e$ & 6 & 3.9 $\times10^{13}$&1.3$\times10^{21}$& 3.1$\times10^{-8}$ \\
-15.5$^f$ & 6 & 6.4 $\times10^{13}$ &2.9$\times10^{21}$& 2.2$\times10^{-8}$ \\
-20.0$^f$ & 5 & 6.5$\times10^{13}$ &4.3$\times10^{21}$& 1.5$\times10^{-8}$ \\
-25.0$^f$ & 6 & 9.7 $\times10^{13}$&4.3$\times10^{21}$& 2.2$\times10^{-8}$  \\
-29.5$^f$ & 4 & 2.9 $\times10^{13}$&2.7$\times10^{21}$& 1.1$\times10^{-8}$  \\
-35.5$^f$ & 5 & 2.3 $\times10^{13}$&1.2$\times10^{21}$& 1.9$\times10^{-8}$ \\
-41.3$^f$ & 4 & 6.9 $\times10^{13}$&1.3$\times10^{21}$& 5.3$\times10^{-8}$  \\
-45.7$^f$ & 4 & 5.1 $\times10^{13}$&4.4$\times10^{21}$& 1.2$\times10^{-8}$  \\
-49.0$^f$ & 3 & 4.6 $\times10^{13}$&2.3$\times10^{21}$& 2.0$\times10^{-8}$  \\
-53.5$^f$ & 6 & 2.9 $\times10^{13}$ &8.0$\times10^{20}$& 3.6$\times10^{-8}$  \\
-58.7$^e$ & 4 & 6.4$\times10^{13}$ &2.7$\times10^{21}$& 2.4$\times10^{-8}$  \\
-69.8$^e$ & 7 & 4.7 $\times10^{13}$&2.0$\times10^{21}$ &2.3$\times10^{-8}$  \\
-76.0$^e$ & 4 & 4.1 $\times10^{13}$&2.1$\times10^{21}$ & 2.0$\times10^{-8}$ \\
-80.1$^e$ & 4 & 2.4$\times10^{13}$ &9.6$\times10^{20}$ & 2.5$\times10^{-8}$  \\
-84.3$^e$ & 5 & 5.7$\times10^{13}$ &2.1$\times10^{21}$ & 2.8$\times10^{-8}$ \\
-88.5$^e$ &5 & 5.1$\times10^{13}$ &1.8$\times10^{21}$ & 2.9$\times10^{-8}$ \\
-93.0$^e$ & 5 & 5.6 $\times10^{13}$&2.4$\times10^{21}$ & 2.3$\times10^{-8}$ \\
-97.5$^e$ & 5 & 4.6$\times10^{13}$ &2.4$\times10^{21}$ & 1.9$\times10^{-8}$ \\
-101.8$^e$ & 4 & 5.9$\times10^{13}$ &4.6$\times10^{21}$ & 1.3$\times10^{-8}$ \\
-106.2$^e$ & 5 & 4.1 $\times10^{13}$ &2.4$\times10^{21}$ & 1.7$\times10^{-8}$\\
-111.6$^e$ &5 & 2.3$\times10^{13}$ &2.2$\times10^{21}$ & 1.1$\times10^{-8}$ \\
-116.0$^e$ &6 & 2.4 $\times10^{13}$&1.7$\times10^{21}$ & 1.4$\times10^{-8}$ \\
\hline
 \end{tabular}
 \end{center}
 {\it a,}{ The H$_{2}$ column densities are decoded from HCO$^{+}$ column densities (Menten et al. 2010). }
     {\it b,}{ Sgr~B2(M); }
     {\it c,}{Scutum arm; } {\it d,}{ Sagittarius arm; } {\it e,}{Galactic center; } {\it f,}{ Norma arm (3 and 4 kpc arms).}
\end{table}

The column densities range from 2.3$\times$10$^{13}$
to 5.2$\times$10$^{14}$~cm$^{-2}$. The Sgr~B2 cloud ($\sim$64
km~s$^{-1}$) has the largest line width and highest column density. Most of
the components have column densities of a few times
10$^{13}$~cm$^{-2}$. The mean column density is
5.4$\times$10$^{13}$~cm$^{-2}$, excluding Sgr~B2(M). This is similar to the \textit{Herschel}/HIFI observations of CH towards the massive star-forming region NGC6334 (Van der Wiel et al. 2010), in which the NGC6334 envelope has the highest column density of 1.7$\times$10$^{14}$~cm$^{-2}$,  while foreground clouds have a column density of several 10$^{13}$~cm$^{-2}$. The CH column
densities in our observations are similar to those in diffuse
clouds and dark nebulae (e.g., Federman 1982;
Liszt \& Lucas 2002; Mattila 1986). Each spiral
arm cloud harbors several velocity
components with different line widths and column densities. The velocity
 structure in each spiral arm cloud suggests that the clouds are not uniform and have internal structure.

\section{Discussion}

For absorption lines, both the observed antenna temperature and
column density of a specific molecule are related to optical depth
and excitation temperature.
The optical depths in this work are derived from the observed
line-to-continuum ratio, therefore the uncertainties due to the
absolute flux calibration will not affect the values of the
optical depth. The brightness temperature of the background continuum
is determined from the baseline offset, which is fairly reliable for HIFI (Bergin et al. 2010). The
only assumption is that the absorption completely covers the
continuum. The spiral-arm cloud sizes are approximately 3 pc
corresponding to 150$^{\prime\prime}$ at a distance of 4 kpc
(Greaves \& Williams 1994), which is more than the size of the
Sgr~B2(M) continuum. The same set of parameters in the LTE model
fits both the spectral features near 532.8 and 536.8 GHz well. The
typical optical depths are 0.5, while the largest optical depth is $\sim$1
at 64 km~s$^{-1}$ 
(see Figure
1). Thus, our LTE fitting gives a reliable assessment of the
column densities.

Determining of the fractional abundance of a molecule relative
to H$_{2}$ requires the column densities of both the molecules in
question and H$_{2}$.  A direct measurement of
H$_{2}$ column density is difficult. HCO$^{+}$ has been shown to have
an approximately constant abundance relative to H$_{2}$ in a
 wide range of environments, including diffuse
clouds, low- and high-mass star-forming regions, and compact
extragalactic continuum sources (e.g., Greaves \& Nyman 1996;
Lucas \& Liszt 1996; van Dishoeck et al. 1993); therefore,
HCO$^{+}$ can be taken as a proxy of H$_{2}$. However, HCO$^{+}$ itself is optically thick in most velocity components.
Following Menten et al. (2010), we use the optically thin H$^{13}$CO$^{+}$
column densities from Table 4 of Menten et al. (2010) to determine
the H$_{2}$ column density and fractional abundance of CH relative
to H$_{2}$ for each velocity component. This procedures takes measured $^{12}$C/$^{13}$C ratios
in the various spiral-arm clouds into account and assumes a constant HCO$^{+}$/H$_{2}$ ratio
of 5$\times$10$^{-9}$.  The fractional abundances
of CH relative to H$_{2}$ are listed in the fifth column of Table
2. The derived H$_{2}$ column densities range from
1$\times$10$^{21}$ to 3$\times$10$^{22}$ cm $^{-2}$ with a total column
density of
1.6$\times$10$^{23}$ cm $^{-2}$ except for Sgr~B2(M),
which are roughly consistent with the values derived from
$^{13}$CO in spiral-arm clouds (Greaves \& Nyman 1996; Neufeld et al. 2000). Most of
the gas components have a CH fractional abundance of a few times
10$^{-8}$ with an average abundance of 2.1$\times$10$^{-8}$,
excluding the Sgr~B2(M) cloud with a value of
9.6$\times$10$^{-10}$ because the excitation there is not
straightforward.

\begin{figure}[h]
\centering
\includegraphics[width=9cm,height=7cm]{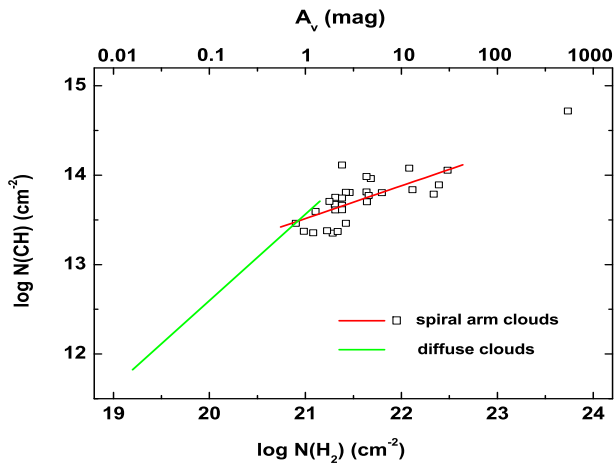}

\caption{Correlation of column densities of CH and H$_{2}$ in a
log-log plot. H$_{2}$ column densities are derived from
H$^{13}$CO$^{+}$ column densities (Menten et al. 2010). The
least-squares fit to the data is indicated by the red line. The
green line shows the slope of 0.97 fitted to the diffuse clouds in
a range of 19.2 $<{\rm log}N({\rm H_{2}})<21.15$ (Sheffer et al.
2008).}\label{}
\end{figure}

Figure 2 presents the relationship between the CH and  H$_{2}$
column densities. A least-squares fit (excluding the Sgr~B2(M)
envelope point) gives a slope of $a=0.38\pm0.07$ and an intercept of
$b=5.81\pm1.64$ (${\rm log}N({\rm CH})=a{\rm log}N({\rm
H_{2}})+b$) with a correlation coefficient of 0.8. The fits to
diffuse clouds (19$<{\rm log}N({\rm H_{2}})<$21.15) give a slope
of $\sim$1 with a correlation coefficient of 0.99 (Federman 1982;
 Sheffer et al. 2008), suggesting that $N({\rm CH})$ correlates
linearly with $N({\rm H_{2}})$. In Figure 2, most of the velocity
components in our observations have higher H$_{2}$ column
densities compared to those of diffuse clouds (Sheffer et al.
2008), and the linear relationship breaks down with N$({\rm H_{2}}) > 10^{21}$ cm$^{-1}$, i.e.\ at visual extinctions beyond 1$^m$, where has a flatter slope of 0.38. The observations of CH and H$_{2}$CO towards a few dense molecular clouds associated with
massive star formation regions (Genzel et al. 1979) have shown
that $N({\rm CH})/N({\rm H_{2}CO})$ decreases with increasing
H$_{2}$CO column density, suggesting that CH abundance is lower in dense
clouds. The relatively flat slope at higher H$_{2}$ column densities
and corresponding lower fractional abundances in our observations corraborate the evidence
that CH abundance is lower in the inner part of the clouds and predominantly traces
lower density environments.  As a word of caution, we note that this hinges on the assumption of constant abundance of HCO$^+$ and has to be tested with other tracers for H$_2$ column densities.

Previous observations have also shown that the CH column density in photodissociation regions (PDR) is not consistent with the linear relationship (Sheffer et al. 2008). Our results appear to agree with the predictions of both turbulent dissipation regions (TDR) and PDR models (see Fig. 8 of Godard, Falgarone \& Pineau des For\^ets 2009) with an H$_{2}$ volume density less than 200~cm$^{-3}$. The chemical network suggests that CH, CH$^{+}$, CH$_{2}$, C, CH$_{3}^{+}$, CH$_{2}^{+}$,  C$^{+}$, C$_{2}{+}$, etc.\ are tightly related by chemistry (Godard, Falgarone \& Pineau des For\^ets 2009). Comparing the column densities of all the hydrides and molecular ions in the absorption line clouds, as provided by HIFI, will enable us to investigate the relative contributions of photon dominated and turbulence dominated chemistry, and distinguish PDR and TDR models.
\begin{acknowledgements}

HIFI has been designed and built by a consortium of institutes and
university departments from across Europe, Canada, and the United
States under the leadership of SRON Netherlands Institute for
Space Research, Groningen, The Netherlands, and with major
contributions from Germany, France, and the US. Consortium members
are: Canada: CSA, U.Waterloo; France: CESR, LAB, LERMA,  IRAM;
Germany: KOSMA, MPIfR, MPS; Ireland, NUI Maynooth; Italy: ASI,
IFSI-INAF, Osservatorio Astrofisico di Arcetri- INAF; Netherlands:
SRON, TUD; Poland: CAMK, CBK; Spain: Observatorio Astron洋ico
Nacional (IGN), Centro de Astrobiolog (CSIC-INTA). Sweden:
Chalmers University of Technology - MC2, RSS \& GARD; Onsala Space
Observatory; Swedish National Space Board, Stockholm University -
Stockholm Observatory; Switzerland: ETH Zurich, FHNW; USA:
Caltech, JPL, NHSC.

\end{acknowledgements}

\end{document}